\documentclass[a4paper,english,aps,prb,twocolumn]{revtex4}
\usepackage[T1]{fontenc}
\usepackage[latin9]{inputenc}
\usepackage{amstext}
\usepackage{graphicx}

\makeatletter
\@ifundefined{textcolor}{}
{%
 \definecolor{BLACK}{gray}{0}
 \definecolor{WHITE}{gray}{1}
 \definecolor{RED}{rgb}{1,0,0}
 \definecolor{GREEN}{rgb}{0,1,0}
 \definecolor{BLUE}{rgb}{0,0,1}
 \definecolor{CYAN}{cmyk}{1,0,0,0}
 \definecolor{MAGENTA}{cmyk}{0,1,0,0}
 \definecolor{YELLOW}{cmyk}{0,0,1,0}
 }

\makeatletter

\usepackage{dcolumn}

\usepackage{bm}

\makeatother

\makeatother

\usepackage{babel}

\begin{document}
\widetext
\title{In-plane magnetic penetration depth in NbS$_{2}$}
\author{P. Diener$^{1}$, M. Leroux$^1$, L. Cario$^{2}$, T. Klein$^{1,3}$, P. Rodière$^{1}$ }
\affiliation{$^{1}$ Institut Néel, CNRS/UJF 25 rue des Martyrs BP 166 38042 Grenoble
cedex 9, FRANCE}
\affiliation{$^{2}$ Institut des Matériaux Jean Rouxel (IMN), Université de Nantes,
CNRS, 2 rue de la Houssinière, BP3229, 44322 Nantes, France}
\affiliation{$^{3}$ Institut Universitaire de France and Universit\'e J.Fourier-Grenoble 1, France}

\begin{abstract}
We report on the temperature dependence of the in plane magnetic penetration depth ($\lambda_{ab}$) and first penetration field ($H_f\propto 1/\lambda_{ab}^2(T)$ for $H\|c$) in 2H-NbS$_2$ single crystals. An exponential temperature dependence is clearly observed in $\lambda_{ab}$(T) at low temperature, signing the presence of a fully open superconducting gap.  This compound is the only superconducting 2H-dichalcogenide which does not develop a charge density wave (CDW). However as previously observed in 2H-NbSe$_2$, this gap ( $\Delta_{1}=1.1\, k_{B}T_{c}$) is significantly smaller than the standard BCS weak coupling value. At higher temperature, a larger gap ($\Delta_{2}=1.8\, k_{B}T_{c}$) has to be introduced to describe the data which are compatible with a two gap model. The superconducting gaps are hence very similar in NbS$_2$ and NbSe$_2$ and we show here that both of them open in the strongly coupled Nb tubular sheets independently of the presence of a CDW or not.    \end{abstract}
\maketitle


\section{Introduction}

In numerous strongly correlated systems, superconductivity
appears in the vicinity of a second electronic instability. These superconductors
then often show deviations from the weak coupling BCS theory which raises
the question of the interplay between the two orders\citep{seyfarthPRL08}.
This question is particularly well illustrated in dichalcogenides. Indeed, in
1T-TiSe$_{2}$\citep{Morosan2006,KusmartsevaPRL2009}, 2H-TaS$_{2}$\citep{Sipos08,Wagner2008},
or 2H-NbSe$_{2}$ superconductivity coexists with a charge density
wave (CDW). In 2H-NbSe$_{2}$ the charge density wave develops for instance at $T_{D}=33\,$K
and coexists with superconductivity (SC) below $T_{c}=7.1\,$K.

In parallele, specific heat, tunneling spectroscopy as well as de Haas van Alphen measurements \citep{Garoche1976,Hess_PRL89,Corcoran1994,Sanchez1995,guillamon:134505} showed that NbSe$_2$ is characterized by the coexistence of two superconducting gaps : $\Delta_{1}=1.0\pm0.1k_{B}T_{c}$ which is much smaller than the BCS weak coupling value ($\Delta/k_BT_c \sim 1.7$) and $\Delta_{2}=2.0\pm0.3k_{B}T_{c}$. It was then initially suggested that the CDW might be at the origin of $\Delta_{1}$\citep{Kiss2007,Fletcher2007,CastroNeto01,Borisenko09}. 

However, scanning tunneling spectroscopy (STS) and specific heat measurements clearly indicated the presence of a reduced
superconducting gap (of similar amplitude) in the NbS$_2$ system as well \citep{guillamon_PRL08,Marcenat09} showing that this interpretation does not hold the ground. Indeed,  this iso-electronic compound is particularly interesting to probe the possible influence of the CDW on the superconducting states in dichalcogenides, as it is the only system among the four 2H layered superconducting transition-metal dichalcogenides \citep{Friend_AdvPhys87} MX$_{2}$
(M=Nb,Ta; X=S, Se) which does {\it not} exhibit a CDW transition. Those measurements hence clearly showed that the presence of two superconducting gaps in 2H-dichalcogenides is {\it not} a direct consequence of the presence of a CDW in some of them.

In a first scenario, those two gaps can be considered as being the limiting values of one single anisotropic superconducting gap (without nodes) whereas in an alternative scenario they can be related to different values of the electron phonon coupling constants ($\lambda_{e-ph}$) on different sheets of the Fermi surface (FS) (so called multiband superconductivity \citep{Suhl_1959}). 

Clear evidence for multiband superconductivity were first obtained in MgB$_{2}$ for which $\lambda_{e-ph}$ varies from $\sim 1.2$ for the 2D-$\sigma$ band to $\sim 0.4$ for the 3D-$\pi$ band \citep{Choi2002,Mazin2003}. Since the FS of NbSe$_2$ is composed of four strongly coupled quasi-2D Nb-derived cylinders (bands 17 and 18 with $\lambda_{e-ph}$ ranging from 0.8 to 1.9) and one Se-derived pancake sheet with a much smaller $\lambda_{e-ph}$ value $\sim0.3$ (band 16)\citep{Corcoran1994}, it was tempting to ascribe the small superconducting gap to this reduced $\lambda_{e-ph}$ \citep{Yokoya2001,PhysRevLett.90.117003}. However ARPES measurements \citep{Kiss2007,Borisenko09} suggested that, in NbSe$_2$, both gaps open on the tubular sheets and the situation iseems to be quite different from that of MgB$_2$ as the two gaps could both be associated to the {\it same} (strongly coupled)  FS sheets. However, an experimental confirmation that the tow gaps are also related to the same FS sheets in NbS$_2$ was still lacking.  

In order to address this issue, we have performed in plane penetration depth and penetration field measurements in NbS$_2$ single crystals. Indeed, $\lambda$ is a directional probe and corresponding measurements are only sensitive to supercurrents flowing perpendicularly to the direction of the applied magnetic field. Band structure calculations in NbSe$_2$ have for instance shown that  the tubular sheets contribute to 98\% of the transport current in the basal plane \citep{johannes:205102}, and the present measurements will hence only probe the superconducting properties of those strongly couple sheets.  We show here that the temperature dependence of $\lambda_{ab}$ are remarkably identical in NbSe$_2$ and NbS$_2$. This temperature dependence clearly confirms the presence of a small superconducting gap $\Delta_{1}=1.1\, k_{B}T_{c}$ (being much smaller than the standard BCS weak coupling value) and that, as previously observed in NbSe$_2$, this gap opens on as least one of the Nb sheets. At higher temperature the superfluid density can be deduced from $\lambda(T)$ introducing a calibration factor (see below) which can be accurately obtained combining  $\lambda(T)$ and first penetration field measurements. We then also show that a second gap value is required to describe the temperature dependence of this superfluid density which can be well described by a two gap model introducing gap values very similar to those previously obtained in specific heat measurements.  We hence confirm that the presence of two well defined superconducting gap values, both opening on the Nb tubular sheets in a generic feature of 2H-dichalconenides.

STS and specific heat ($C_p$) measurements clearly showed that NbSe$_2$ and NbS$_2$ systems share the particularity of having both two superconducting gaps but also highlighted some fundamental differences. Indeed, whereas vortices exhibit a clear stellar shape in NbSe$_2$, they appear to be isotropic in NbS$_2$. Moreover, the two systems have very different anisotropies of their upper critical fields increasing from $\sim 3.2$ in NbSe$_2$ \citep{Sanchez1995} to $\sim 8.1$ in NbS$_2$ \citep{Onabe_JPSJ78}. However, the vortex shape and $H_{c2}$ values are determined by the coherence length ($\xi$) which is related to the superconducting gap but also to electronic properties of the normal state (Fermi velocity and possibly electronic mean free path in the so-called dirty limit). On the contrary the temperature dependence of the penetration depth is determined {\it only} by the superconducting gap structure. Those later measurements hence provide very valuable information to understand those differences which will be discussed at the end of the paper.

\section{sample and experiments}

2H-NbS$_{2}$ samples were grown using the vapour transport growth technique
with a large sulphur excess, as described elsewhere \citep{Fisher_IC80}.
Single crystals have a flat hexagonal or triangular shape and
typical dimensions of $150^{2}\times30\,\mu m$. The large flat faces grow 
perpendicular to the c-axis. The 2H crystallographic
structure has been checked on each sample using a four-circle diffractometer.
The bulk critical temperature was checked on several samples of each
batch by ac specific heat \citep{Marcenat09}. All the measured crystals
showed sharp superconducting transitions ($\Delta T_{c}\sim0.3\,$K)
at $T_{c}=6.05\pm0.2\,$K emphasizing the excellent bulk homogeneity
of each crystal. Previous scanning tunnelling microscopy measurements
on single crystals of the same origin confirmed the absence of CDW
down to 0.1 K \citep{guillamon_PRL08}. To avoid a contamination by
oxygen intercalation between layers which would reduce $T_{c}$ and smears
the transition\citep{Onabe_JPSJ78}, samples were kept in vacuum. 
All the measurements were carried out in the first month following 
the growth of the samples.

\begin{figure}[t]
\begin{centering}
\includegraphics[bb=140 140 489 409,clip,width=8cm]{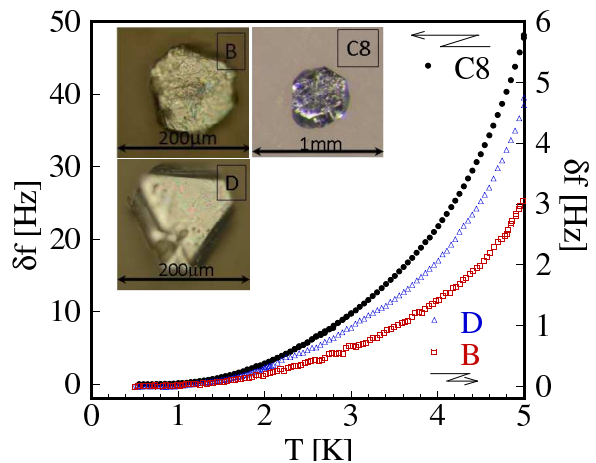}
\end{centering}
\caption{\textit{\small \label{fig:NbS2_reproductibilite}Temperature dependence
of the frequency shift $\delta f$(T), for B (square) D (triangle) (right scale)and
C8 (circle) (left scale)NbS$_{2}$ single crystals. Inset: microscope images of each sample
(note the bigger scale for C8). }}
\end{figure}

A high stability
LC oscillator operating at 14$\,$MHz, driven by a Tunnel Diode \citep{degrift:599,prozorov:4202,Diener_PRB09} has been used to measure the variation of the magnetic penetration depth. The tiny sample, placed in the center of the solenoid coil, acts as a small perturbation to the inductance (L). By changing the temperature of the sample, the magnetic penetration depth variations produce a change of the inductance, leading to a change in the LC resonant frequency $\delta f(T)$.  The amplitude of the AC excitation field is smaller than 1$\mu$T (that is <<$H_{c1}$) to keep the sample in the Meissner state.
This technique enables to measure the variation of $\lambda$ with an excellent sensitivity as our setup can detect relative frequency changes, $\delta f/f_0$ of a few 10$^{-9}$, which corresponds to changes in  $\lambda$ on the order of $1$ \AA   in mm$^3$ samples.
The sample is glued with vacuum grease at the bottom of a sapphire cold finger whose temperature is regulated between $T_{min}$=0.5 K and 10 K whereas the LC oscillator remains at fixed temperature (3K). 
The frequency shift $\delta f$=$f(T)-f(T_{min}$) is then
proportional to the magnetic penetration depth : 
\begin{equation}
\delta\lambda(T)=R\frac{\delta f(T)}{\Delta f_{0}}
\end{equation}
where $\Delta f_0$ is the change induced by the introduction of the superconducting sample in the coil (measured in-situ by extracting the cold finger from the coil at the lowest temperature), and $R$ an effective dimension which depends on the sample geometry. In principle, $R$ can be evaluated by using the formula proposed by Prozorov and al. \citep{prozorov_cal} in the case of cylinders. However, recent measurements in pnictides highlighted the difficulties of evaluating this effective dimension, in particular for magnetic fields perpendicular to the ab-planes (main sample surface) \citep{Klein2010}. Indeed, very different  $\delta\lambda(T)$ were obtained by different groups, most probably due to the roughness of the sample edges which undermines this evaluation.

To overcome this difficulty we have performed first penetration field ($H_f$) measurements using miniature GaAs-based quantum well Hall probe arrays. The remanent local DC field ($B_{rem}(H_a)$) in the sample has been measured after applying a magnetic field $H_a$ and sweeping the field back to zero. In the Meissner state, no vortices penetrate the sample and $B_{rem}$ remains equal to zero up to $H_a$ = $H_f$. A finite $B_{rem}$ value is then obtained for $H_a \geq H_f$ due to vortices which remain pinned in the sample. As $H_f$ is directly proportional to the first critical field $H_{c1} \propto 1/\lambda^2$ , the  normalized in plane superfluid density :
\begin{equation}
\tilde{\rho}_{ab}(T)=\frac{\lambda_0^{2}}{\lambda_{ab}^2(T)}=\frac{1}{(1+\frac{R}{\lambda_0 \Delta f_0}\delta f(T)/)^2}
\end{equation}
 is expected to vary as $H_f(T)/H_f(0)$ ($\lambda_0$ being the zero temperature penetration depth). Even though the uncertainties on the $H_f$ values hinders any precise determination of the gap values,  those measurements can then be used as milestones to obtain the $R/\lambda_0\Delta f_0$ coefficient (see inset of Fig.3).

\section{Evidence for a small gap value in NbS$_2$}

Fig. \ref{fig:NbS2_reproductibilite} shows typical $\delta f(T)$
curves  between $0.08\, T_{c}$ and $0.85\, T_{c}$ for three NbS$_{2}$ single crystals (see inset of 
Fig. \ref{fig:NbS2_reproductibilite}). The magnetic field is applied perpendicularly to the layers, i.e. along the crystallographic
$c$ axis, so that supercurrents are flowing in the layers,
probing the in-plane penetration depth, $\lambda_{ab}$ \citep{Chandrasekhar_93}.  Samples B and D were grown in the same batch, and have, in-plane
dimensions equal to 110$\,\mu m$ and 140$\mu m$, respectively and a thickness
$t = 30 \mu$m (resp. $25 \mu$m). Sample C8 was grown in a second batch
and has a larger size ($430^{2}\times110\,\mu m^{3}$). The frequency
shift is ten times higher for C8 than for other samples due to it
larger volume but all renormalized curves are fully identical underlying the excellent reproducibility of the temperature dependence discussed below.

\begin{figure}[t]
\begin{centering}
\includegraphics[bb=116 288 457 562,clip,width=8cm]{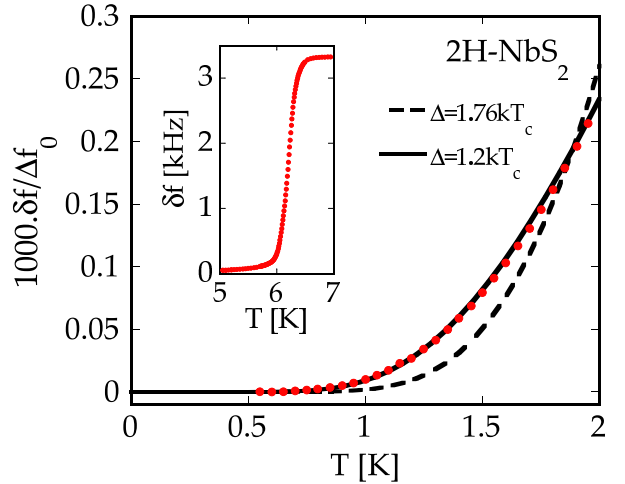}
\par\end{centering}
\caption{\textit{\small \label{fig:NbS2_DL_BT}Temperature dependence of the
rescaled frequency shift $\delta f/\Delta f_0 \propto \Delta \lambda_{ab}/\lambda_0$ in NbS$_{2}$ at T<T$_{c}$/3 (red circles).
Results are fitted with the exponential low temperature approximation
(see text) to determine the smaller gap $\Delta$. An excellent fit
is obtained for $\Delta\approx1.17\, k_{B}T_{c}$. Results can't be
described by an exponential behaviour with the weak coupling BCS value
$\Delta=1.76\, k_{B}T_{c}$. Inset: superconducting transition: T$_{c}$=6.05$\,$K
is determined by the maximum of variation of the frequency shift. }}
\end{figure}

In the local London theory, at low temperature (typically $T< T_c/3$), the temperature dependence of the gap can be neglected and $\lambda(T)$ is given  by :
\begin{equation}
(\lambda(T)-\lambda_0)/\lambda_{0}=\sqrt{\frac{\pi\Delta(0)}{2k_{B}T}}e^{-\Delta(0)/k_{B}T}\label{eq3}
\end{equation}
where $k_B$ is the Boltzmann constant and $\Delta(0)$ the value of the {\it smallest} superconducting 
gap at 0 K. Indeed, due to this exponential dependence only the smallest gap is expected to show up at low temperature.  As shown in Fig.\ref{fig:NbS2_DL_BT} a very poor agreement to the data is obtained if the gap parameter is fixed to the weak coupling BCS gap of 1.76$k_{B}T_{c}$. A much better fit is obtained taking $\Delta/k_{B}=7.2\, $K$\approx1.2\, T_{c}$.  Note that this gap value does not depend on the renormalization constant. The assumption $\lambda(T_{min})\approx\lambda_0$, where T$_{min}$ is the lowest experimental temperature,  is valid here, since an extrapolation of Eq.(3) gives $(\lambda(0.5K)-\lambda_{0})/\lambda_0 <10^{-3}$. Since the superfluid current mainly originates from the Nb sheets for $H\|c$, those measurements hence directly show that this small gap is associated with those sheets in NbS$_2$ as well.

\begin{figure}[t]
\begin{centering}
\includegraphics[bb=100 350 431 630,clip,width=8cm]{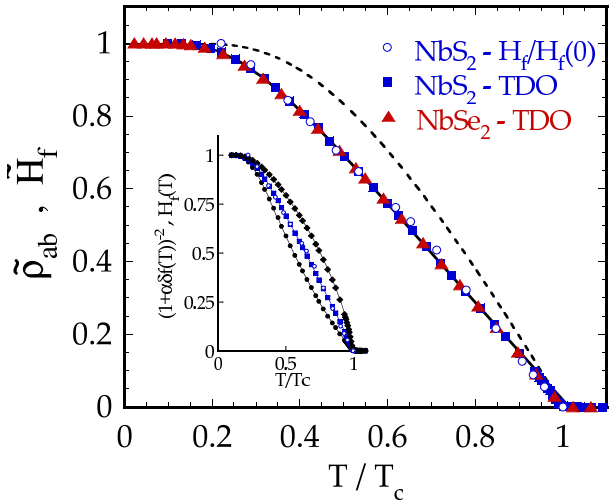}
\par\end{centering}
\caption{\textit{\small \label{fig:NbS2_nS}Normalized superfluid density of NbS$_{2}$ deduced from Tunnel Diode Oscillator measurements (TDO) for $R\lambda_{0}/\Delta f_0$=0.02Hz$^{-1}$ (closed squares), compared to $H_f(T)/H_f(0)$ (open circles) both on sample C8. As shown, the weak coupling BCS normalized superfluid density (dashed line) is not compatible with the data. A very similar temperature dependence was previously obtained in NbSe$_2$ (solid triangles) and both compounds could be well described by a two gap model (thick line, see text for details). Inset: influence of the $R/\lambda_{0}\Delta f_0$ parameter on the shape of the superfluid density, $R/\lambda_{0}\Delta f_0$ being equal to respectively 0.01, 0.02 and 0.03 Hz$^{-1}$ for diamonds, squares and circles.  }}
\end{figure}

In order to get information on the larger gap, it is necessary to perform a detailed analysis of the temperature dependence at higher temperature. The inset of fig.\ref{fig:NbS2_nS} displays the temperature dependence of the-superfluid density 
$\tilde{\rho}_{ab}(T)$ up to $T_c$ deduced from Eq.(2) for different $R/\lambda_0\Delta f_0$ values compared to $H_f/H_f(0)$. Note that if a small change in $R/\lambda_0\Delta f_0$ will modify strongly the normalized superfluid density, the curvature of $H_f$(T) is utterly independent of $\lambda_0$. $H_f$(T) can hence be used to determine $R/\lambda_0\Delta f_0$ with good accuracy and, as shown, a good agreement between $H_f$ and the superfluid density is obtained taking $R/\lambda_0\Delta f_0=0.02Hz^{-1}$. Clear deviations from the BCS weak coupling theoretical curve (dashed line) is visible on the entire temperature range. In particular, the BCS values exceed the experimental data at low temperature due to the small value of the gap.

In the multigap model,  $\tilde{\rho}$ is phenomenologically assumed to 
be the sum of two independent superfluid densities 
\begin{equation}
\tilde{\rho}=\omega \tilde{\rho_1}+(1-\omega) \tilde{\rho_2}
\end{equation}

where $\omega$ is the weight of band 1 and $\tilde{\rho_i}$ are given by

\begin{equation}
\tilde{\rho_{i}}(T)=1-\int\frac{\partial f}{\partial E}\frac{E}{\sqrt{E^{2}-\Delta_{i}^{2}(T)}}
\end{equation}

The best fit to the data using Eq.(4) and Eq.(5) (solid line in  Fig.\ref{fig:NbS2_nS}) is obtained with $\Delta_{1}=1.05\pm0.05\times k_{B}T_{c}$ and
$\Delta_{2}=1.8\pm0.1\times k_{B}T_{c}$ with $\omega \sim 0.5$.
Those values are in very good agreement with those deduced from tunneling spectroscopy measurements which revealed a distribution of gaps having
two main values at 0.53\,meV=6.15\,K$\approx1\, k_{B}T_{c}$ and
0.98\,meV=10.8\,K$\approx1.8\, k_{B}T_{c}$ \citep{guillamon_PRL08}.
Moreover, the temperature dependence of the specific heat measured on sample C8 can be well described taking $\Delta_1 \sim 1.05\times k_{B}T_{c}$, $\Delta_2 \sim 2.1\times k_{B}T_{c}$ and $\omega \sim 0.6$ \citep{Marcenat09}. 
Even if this two gap model well reproduces the temperature dependence of $\tilde{\rho_{i}}$ it is worth noting that this temperature dependence is also compatible with a single anisotropic in-plane superconducting gap. The simplest model which respects the symmetry of the crystal has an in-plane 6-fold shape, $\Delta(\phi)=\Delta_1+\frac{\Delta_2-\Delta_1}{2}(1+cos(6\phi))$ where the extremum of the superconducting gap define two characteristics energy scales. With this model, we find $\Delta_1\sim 0.9(1)\times k_{B}T_{c}$ and $\Delta_2\sim 1.9(3)\times k_{B}T_{c}$.

\section{Comparison between N\textsc{b}S$_2$ and N\textsc{b}S\textsc{e}$_2$}

We have also reported in Fig.\ref{fig:NbS2_nS} the temperature dependence of $\tilde{\rho}_{ab}$ deduced from penetration depth measurements in 2H-NbSe$_{2}$ \citep{Fletcher2007}. Since the critical temperatures are slightly different in the two compounds, the data have been plotted as a function of $T/T_c$ with $T_{c} = 6.05\,$K and $7.1\,$K for NbS$_{2}$ and NbSe$_{2}$, respectively. In both case, the presence of two characteristic gap values are necessary to fit the temperature dependence of the in-plane superfluid density.
We hence show that in both compounds the exotic superconducting gap structure is related to the Nb tubular sheets and that, even if the charge density wave is perturbing those sheets in NbSe$_2$ (but not in NbS$_2$ which does not exhibit any CDW), this CDW does not affect the superconducting gap structure. Our results hence contradict the ARPES measurements on NbSe$_2$ \citep{Kiss2007} which suggested a strong correlation between the gap structure and the CDW formation. The presence of an identical exotic gap structure in two different compounds (even of the same family) is particularly rare which highlights the robustness of the underlying mechanism at its origin.
It's worth noting that there exists a $R/\lambda_0\Delta f_0$  value ($\sim 0.02Hz^{-1}$, see above)  for which the temperature dependence obtained on both compounds are very  similar on the whole temperature range. 

Note that this strong similitude in the gap structure is not in contradiction with the fact that the vortex cores have an isotropic shape in NbS$_2$ \citep{guillamon_PRL08}, and  6-fold shape in NbSe$_2$ \citep{Hess_PRL90,Guillamon2008}. Indeed, in this later case, the particular shape could be due to an anisotropic Fermi velocity \citep{Gygi_PRL90} caused by the presence of the Charge Density Wave.

Finally, It is interesting to note that even with similar superconducting gaps, it is possible to understand the large difference of the anisotropy of the upper critical field in the two systems. In the case of NbSe$_2$, the anisotropy of the upper critical field can be understood by the anisotropy of the two set of Nb-bands. Indeed, the coherence length related to the pancake band is larger due to a large renormalised Fermi velocity and a small gap \cite{Corcoran1994}. According to band structure calculations, the anisotropy of the Fermi velocities ($v_F^{ab}/v_F^c$) of the two tubular sheets are on the order of $2$ (Band 17 \cite{Remarque}) and $10$ (Band 18) in NbSe$_2$, none of which corresponding to the experimental value of the anisotropy of the upper critical field ($\sim 3.2$)\cite{johannes:205102}. However, $\mu_0H_{c2}^{\|c}(0)=\Phi_0/2\pi\xi_{ab}^2$ is defined by the band having the smallest $\xi_{ab}$ value. All bands have quite similar in plane Fermi velocities but if one assumes that $\Delta_2$ is associated with band 17 and $\Delta_1$ to band 18, on gets $\xi_{ab,17} \sim (\Delta_1/\Delta_2)\xi_{ab,18} \sim  \xi_{ab,18}/1.7$ and $H_{c2}^{\|c}$ will be defined by band 17. Moreover, for $\mu_0H_{c2}^{\|ab}(0)=\Phi_0/2\pi\xi_{ab}\xi_c$, and  the very small value of c-axis Fermi velocity of band 18, will imply that $H_{c2}^{\|ab}$ will be defined by this latter band. One hence gets $H_{c2}^{\|ab}/ H_{c2}^{\|c} = (v_F^{ab,17}.v_F^{ab,17}/v_F^{ab,18}.v_F^{c,18})\times(\Delta_1/\Delta_2)^2 \sim 10/(1.7)^2 \sim 3$ in good agreement with the experimental value. Note that this situation strongly differs from the one observed in MgB$_2$ in which the anisotropy of $H_{c2}$ is defined only by the $\sigma-$band (for $T << T_c$). As NbSe$_2$ and NbS$_2$ have similar $H_{c2}^{\|ab}$ values, one can conclude that band 18 is not affected by the chemical substitution and the increase of the $H_{c2}$ anisotropy to $\sim 8.1$ in NbS$_2$ is a hence due to an increase of $v_F^{ab,17}$ but this change in the Fermi velocity does not affect the superconducting gap structure.

The presence of two characteristics gap values appears to be a robust characteristic feature in 2H dichalcogenides, but its origin is still open to questions. With the present results, the influence of the CDW also has to be excluded, underlying the necessity to propose new clues for the origin of the reduced superconducting gap observed in both compounds.


\section{Conclusion}

To conclude, the temperature dependence of the in-plane magnetic penetration
depth of single crystals of 2H-NbS$_{2}$ has been measured, and compared
to previous measurements on 2H-NbSe$_{2}$. The temperature dependence shows a similar curvature in both compounds evidencing a similar reduced superconducting gap structure in the sheets of the Fermi Surface where the e-ph coupling
constant are the largest. The CDW,
present only in NbSe$_{2},$ is not at the origin of the reduced superconducting
gap. In the light of the discovery of a particular acoustic phonon
mode observed in lead or Niobium \citep{Aynajian_08}, a more accurate
knowledge of the phonon dispersion in NbS$_{2}$ could shed insight
into the origins of the reduced superconducting gap in the dichalcogenides.

P.R. thanks R. and B. Thomas for helpful discussions.  
\bibliographystyle{apsrev}. TK is most obliged to V. Mosser of ITRON, Montrouge,
and M.Konczykowski from the Laboratoire des Solides Irradi\'es, Palaiseau for the development of the Hall sensors used in this study
\bibliography{NbS2}

\end{document}